\documentclass[aps, prl, twocolumn, superscriptaddress]{revtex4-1}

\usepackage{amsmath}	
\usepackage{amsthm}		
\usepackage{amssymb}	
\usepackage{eufrak}
\usepackage{datetime}
\usepackage{graphicx}
\usepackage{color}
\usepackage{verbatim}
\usepackage{bm}
\usepackage{float}
\usepackage{bbold}

\usepackage{dsfont}

\usepackage[colorlinks=true, citecolor=midblue, linkcolor=midblue, urlcolor=midblue]{hyperref}

\usepackage{setspace}

\settimeformat{ampmtime}

\definecolor{grey}{rgb}{0.4,0.4,0.4}
\definecolor{dullmagenta}{rgb}{0.4,0,0.4}
\definecolor{darkblue}{rgb}{0,0,0.4}
\definecolor{midblue}{rgb}{0,0,0.5}
\definecolor{midred}{rgb}{0.5,0,0}
\definecolor{orange}{rgb}{1,0.5,0}
\definecolor{lightbrown}{rgb}{0.75,0.5,0.25}
\definecolor{tan}{cmyk}{0.14,0.42,0.56,0}
\definecolor{djunglegreen}{cmyk}{0.99,0,0.52,0}
\definecolor{lightgreen}{rgb}{0,1,0}
\definecolor{olivegreen}{cmyk}{0.64,0,0.95,0.40}
\definecolor{midgreen}{rgb}{0.0,0.675,0.0}
\definecolor{darkgreen}{rgb}{0,0.5,0}

\newcommand{\vs}{\vspace}

\renewcommand{\.}{\hspace{0.5mm}}

\newcommand{\Grm}{\ensuremath{\mathrm{G}}}

\newcommand{\Prm}{\ensuremath{\mathrm{P}}}

\newcommand{\Urm}{\ensuremath{\mathrm{U}}}

\newcommand{\drm}{\ensuremath{\mathrm{d}}}
\newcommand{\erm}{\ensuremath{\mathrm{e}}}

\newcommand{\grm}{\ensuremath{\mathrm{g}}}

\newcommand{\vrm}{\ensuremath{\mathrm{v}}}
\newcommand{\wrm}{\ensuremath{\mathrm{w}}}

\newcommand{\Lcal}{\ensuremath{\mathcal{L}}}

\renewcommand{\d}{\ensuremath{\mathrm{d}}}

\newcommand{\cf}{c.f.}

\settimeformat{ampmtime}

\usepackage{float}

\begin{document}

\title{Vortices in Black Holes}

\author{Gia Dvali}
\affiliation{
	Arnold Sommerfeld Center,
	Ludwig-Maximilians-Universit{\"a}t,
	Theresienstra{\ss}e 37,
	80333 M{\"u}nchen,
	Germany}
\affiliation{
	Max-Planck-Institut f{\"u}r Physik,
	F{\"o}hringer Ring 6,
	80805 M{\"u}nchen,
	Germany}

\author{Florian K{\"u}hnel}
\email{Florian.Kuehnel@physik.uni-muenchen.de}
\affiliation{
	Arnold Sommerfeld Center,
	Ludwig-Maximilians-Universit{\"a}t,
	Theresienstra{\ss}e 37,
	80333 M{\"u}nchen,
	Germany}

\author{Michael Zantedeschi}
\email{Michael.Zantedeschi1@physik.uni-muenchen.de}
\affiliation{
	Arnold Sommerfeld Center,
	Ludwig-Maximilians-Universit{\"a}t,
	Theresienstra{\ss}e 37,
	80333 M{\"u}nchen,
	Germany}
\affiliation{
	Max-Planck-Institut f{\"u}r Physik,
	F{\"o}hringer Ring 6,
	80805 M{\"u}nchen,
	Germany}

\date{\formatdate{\day}{\month}{\year}, \currenttime}

\begin{abstract}
We argue that black holes admit vortex structure. This is based both on a graviton-condensate description of a black hole as well as on a correspondence between black holes and generic objects with maximal entropy compatible with unitarity, so-called saturons. We show that due to vorticity, a $Q$-ball-type saturon of a calculable renormalizable theory obeys the same extremality bound on the spin as the black hole. Correspondingly, a black hole with extremal spin emerges as a graviton condensate with vorticity. This offers a topological explanation for the stability of extremal black holes against Hawking evaporation. Next, we show that in the presence of mobile charges, the global vortex traps a magnetic flux of the gauge field. This can have macroscopically-observable consequences. For instance, the most powerful jets observed in active galactic nuclei can potentially be accounted for. As a signature, such emissions can occur even without a magnetized accretion disk surrounding the black hole. The flux entrapment can provide an observational window to various hidden sectors, such as millicharged dark matter.
\end{abstract}

\maketitle

{\textsl{Introduction}\;---\;}The microscopic structure of black holes remains to be understood. One of the main obstacles is the lack of experimental probes of black hole quantum properties. In this light, it is very important to identify and explore those microscopic theories that lead to macroscopically-observable phenomena. In this note we shall create awareness about one such phenomenon: vorticity. Our proposal is based on two lines of thought. 

On one hand, there are physical indications \cite{Dvali:2011aa, Dvali:2012en} that at the quantum level a black hole of radius $R$ represents a condensate of ``soft" gravitons, i.e., gravitons of characteristic wavelength $\sim R$ and frequency
\begin{equation}
	\label{eq:frequency} 
	\omega
		\sim
					1 / R
					\, . 
\end{equation} 
The defining property of a black hole is that the graviton condensate is at the critical point of saturation, also referred to as the point of ``maximal packing". At this point, the occupation number of gravitons $N_{\rm gr}$, their quantum gravitational coupling (we shall work in $3 + 1$ dimensions), 
\begin{equation}
	\label{eq:AlphaGR} 
	\alpha_{\rm gr} 
		\equiv 
					1 / ( R\mspace{1.5mu}M_{\Prm} )^{2}
					\, ,~~
	M_{\Prm}
		\equiv
					{\rm Planck~mass}
					\, , 
\end{equation} 
and entropy $S$, satisfy the relation
\begin{equation}
	\label{eq:Criticality} 
	S
		=
					N_{\rm gr}
		=
					\frac{ 1 }{ \alpha_{\rm gr} } 
					\, . 
\end{equation} 
Taking into account Eq.~\eqref{eq:AlphaGR}, it is clear that this equation, reproduces the Bekenstein-Hawking black hole entropy. 

On the other hand, it has been argued recently \cite{Dvali:2020wqi} that the expression \eqref{eq:Criticality} is not specific to black holes or gravity. Rather, it represents a particular manifestation of the following universal upper bound, imposed by unitarity, on the microstate entropy $S$ of a generic self-sustained bound state of size $R$,
\newpage

\begin{align}
	&
	\notag
	\\[-13mm]
 S
		&=
					\frac{ 1 }{ \alpha }
		=
					R^{2} f^{2}
					\, ,
					\vs{-1mm}
					\label{eq:Sbound} 
\end{align} 
where $\alpha$ is the running coupling of the theory evaluated at the scale $R$ and $f$ is the decay constant of Goldstone modes of symmetries spontaneously broken by the bound state. Furthermore, it was argued that there exists a large class of objects saturating this bound and that such objects, so-called saturons, exhibit close similarities with black holes. This proposal has been verified on multiple examples \cite{Dvali:2019jjw, Dvali:2019ulr, Dvali:2020wqi, Dvali:2021rlf, Dvali:2021tez}. The correspondence between black holes and generic saturated systems opens up the possibility of using saturons in calculable theories as laboratories for understanding the well-established black hole properties and for predicting new ones. 
 
The goal of the present paper is to provide one more link in this correspondence. Our main message is that black holes and other saturons naturally support a vortex structure. This imposes an upper bound on the saturon spin very similar to the extremality bound on a spinning black hole. This offers a microscopic explanation of black hole's maximal spin in terms of the vorticity of the graviton condensate. 

The vortex structure, when interacting with a neutral plasma of particles charged under some gauge symmetry, such as electromagnetism, necessarily traps a magnetic flux in it. This trapping can have some macroscopically-observable consequences. It can also provide an observational window in various hidden sectors, such as millicharged dark matter.
 
We believe that, due to the universality of saturons, the vorticity property in black holes can be understood without entering into the technicalities of quantum gravitational computations. In short, at the level of our presentation, the vorticity in black holes represents a conjecture supported by the evidence gathered from calculable saturated systems. 
\newpage

We shall explain the main concepts on a prototype SU$( N )$ model \cite{Dvali:2020wqi} which has been shown by Ref.~\cite{Dvali:2021tez} to support the black-hole-like saturated bound-states. These bound-states can be viewed as a special version of non-topological solitons, or $Q$-balls \cite{Coleman:1985ki, Lee:1991ax} (for some implications, see Refs.~\cite{Kusenko:1997ad, Dvali:1997qv}). An important new feature that makes them black-hole-like, is the maximal degeneracy of microstates which satisfies Eq.~\eqref{eq:Criticality}.\\[-2mm]

{\textsl{Prototype Model}\;---\;}To set the stage, let us consider a theory \cite{Dvali:2020wqi, Dvali:2021tez} of a scalar field $\phi$ in the adjoint representation of global SU$( N )$ symmetry. The Lagrangian density in obvious matrix notations has the following form,
\begin{align}
	\label{eq:Model}
	\Lcal
		&=
					\dfrac{ 1 }{ 2 }\.
					\text{tr}\!
					\left(
						\partial_{\mu} \phi
					\right)\!
					\left(
						\partial^{\mu} \phi
					\right)
					-
					\dfrac{\alpha}{2}\.\text{tr}\!
					\left(
						f \phi
						-
						\phi^{2}
						+
						\dfrac{ \mathds{1} }{ N }\.
						\text{tr}\phi^{2}
					\right)^{\!2}
					\, .
\end{align}
Here, $\alpha$ is a dimensionless coupling and $f$ is the scale. It is very important to keep in mind that $\alpha$ and $N$ obey the unitarity constraint \cite{Dvali:2020wqi}
\begin{equation} 
	\label{eq:Uni} 
	\alpha\mspace{1mu}N\mspace{-1mu}
		\lesssim
					1
					\, . 
\end{equation} 
The correspondence with a macroscopic black hole pushes us to consider the extremely large values of $N$ (for instance, considering a solar-mass black hole leads to $N \sim 10^{77}$). This gives us the benefit of using the power of $1 / N$-expansion. Throughout the paper, we shall work in the leading order approximation in $1 / N$. 
 
The theory \eqref{eq:Model} has multiple degenerate vacua with different patterns of spontaneous symmetry breaking. We shall focus on the SU$( N )$-symmetric vacuum and the one with the symmetry broken to SU$( N - 1 ) \times \Urm( 1 )$. In the former vacuum, the theory exhibits a mass gap $m^{2} = \alpha f^{2}$. In the latter, there exist $\simeq 2 N$ species of massless Goldstone modes. The two vacua are separated by a domain wall. A planar static wall has a tension (energy per unit length) equal $m^{3} / 6 \alpha$. The thickness of the wall is $\delta_{\wrm} = 1 / m$. 

Following Ref.~\cite{Dvali:2021tez}, we shall be interested in the field configurations described by the ansatz
\begin{equation}
	\label{eq:rotation}
	\phi
		=
					U^{\dagger} \Phi\mspace{2mu} U
					\, ,
\end{equation}
where 
\begin{equation}
	\label{eq:phi} 
	\Phi
		=
					\dfrac{ \rho( x ) }{ N }\.
					\text{diag}
					\big[
						( N - 1 ), - 1, \dotsc, - 1
					\big]
					\, ,
\end{equation}
and
\begin{equation}
	U
		=
					\exp\mspace{-3mu}
					\big[
						i\.\theta( x )\.T
					\big]
					\, .
\end{equation}
Here $T$ is one of the generators broken by Eq.~\eqref{eq:phi} and $\theta( x )$ is the corresponding Goldstone mode. This ansatz gives the effective Lagrangian of two degrees of freedom ($1 / N$ effects absorbed in rescalings)
\begin{equation}
	\label{eq:Lag}
	\Lcal
		=
					\frac{ 1 }{ 2 }\.
					\big[
						\partial_{\mu}\rho\.
						\partial^{\mu}\rho
						+
						\rho^{2}\.
						\partial_{\mu}\theta\.
						\partial^{\mu}\theta
						-
						\alpha\.\rho^{2}\mspace{1.5mu}
						(\rho
							-
							f
						)^{2}
					\big]
					.
\end{equation}
In this effective theory, we shall be interested in the state that represents a hybrid of the following two solutions. 

The first solution \cite{Dvali:2021tez}, which serves as the basis for the black hole prototype, is a stationary spherical bubble of SU$( N - 1 ) \times \Urm( 1 )$ vacuum embedded in SU$( N )$-invariant asymptotic space. This solution is described by the ansatz,
\begin{equation}
	\label{eq:ansatz0}
	\rho
		=
					\rho( r )
					\, , ~ 
	\theta
		=
					\omega\.t
					\, . 
\end{equation}
Correspondingly, the only non-trivial equation is
\begin{equation}
	\label{eq:radial}
	\drm_{r}^{2}\rho
	+
	\dfrac{ 2 }{ r }\.\drm_{r}\rho
	+ 
	\rho
	\left[
		\omega^{2}
		-
		\alpha\.
		( \rho - f )\.
		( 2\mspace{1.5mu}\rho - f)
	\right]
		=
					0
					\, .
\end{equation}
This has a solution that interpolates from $\rho( 0 ) = f$ to $\rho( \infty ) = 0$.
 
The bubble is stable because of the conserved charge, equal to the occupation number of the Goldstone modes, $N_{\Grm}$. In the thin-wall approximation, 
\begin{equation}
	\label{eq:thinwall}
	\omega^{2}
		\ll
					\alpha\.f^{2}
		=
					m^{2}
					\, ,
\end{equation}
the charge is given by,
\begin{equation}
	\label{eq:Charge}
	N_{\Grm}
		=
					2\pi\mspace{2mu}\omega
					\int \drm r\;
					r^{2}\.\rho^{2}( r )
		\simeq 
					\dfrac{ 2\pi }{ 3\alpha }\.
					m^{2}\.
					\omega\mspace{1mu}R^{3}
					\, .
\end{equation}
Correspondingly, in this regime the energy of a stationary bubble, its radius and charge satisfy, 
\begin{equation}
	\label{eq:ERNG} 
	M_{\rm Bub} =
					\dfrac{ \omega }{ \alpha }\.
					\frac{ m^{5} }{ \omega^{5} }\!
					\left(
						\frac{ 40\pi }{ 81 }
					\right )\!
					,~ 
	R =
					\dfrac{ 2 }{ 3 }\.
					\frac{ m }{ \omega^{2} }
					\. ,~
	N_{\Grm}
		=
					\dfrac{ 1 }{ \alpha }\.
					\dfrac{ m^{5} }{ \omega^{5} }\!
					\left(
						\dfrac{ 16\pi }{ 81 }
					\right)_{_{_{_{_{_{_{}}}}}}}
				 \mspace{-8mu}.
\end{equation} 
All the above relations, up to order one factors, hold also in the thick-wall regime, in which we have 
\begin{equation}
	\label{eq:thickwall}
	\omega 
		\sim 
					m
		\sim 
					\frac{ 1 }{ R }
					\, . 
 \end{equation} 
It was observed \cite{Dvali:2021tez} that in this (i.e., thick-wall) regime, the bubble exhibits a striking correspondence with a black hole when the system saturates the upper bound on unitarity \eqref{eq:Uni}. At this point, the characteristics of the saturon bubble of the SU$( N )$ theory become isomorphic to an equal-radius black hole of gravity, under the mapping $f \rightarrow M_{\Prm}$ 
In particular, the mass, entropy and radius of the saturon bubble satisfy, 
\begin{equation}
	\label{eq:mass}
	S
		\sim
					\frac{ 1 }{ \alpha }
		\sim
					( R\mspace{1mu}f )^{2}
		\sim
					\frac{ M_{\rm Bub}^{2} }{ f^{2} }
					\, . 
\end{equation} 
The source of the bubble's microstate entropy, which matches the Bekenstein-Hawking entropy of the corresponding black hole, is the exponential degeneracy of the Goldstone vacuum in the bubble interior. From now on, the thick-wall regime will be the focus of this work.

In order to establish the basis for vorticity, let us notice that the Lagrangian \eqref{eq:Lag} admits a second, topologically non-trivial, configuration in form of a global vortex line \cite{Vilenkin:2000jqa}. In polar coordinates $(r, \varphi, z)$ it is described by the ansatz
\vs{-1mm}
\begin{equation}
	\label{eq:ansatz01}
	\rho
		=
					\rho( r )
					\, ,~
	\theta
		=
					n
					\.\phi
					\, , 
\end{equation} 
where $\rho( r )$ is interpolating between $\rho( 0 ) = 0$ and $\rho( \infty ) = f$. The existence of the solution is guaranteed by the topologically non-trivial boundary condition with winding number $n$. The core of the vortex has a size $\delta_{\vrm} \sim 1 / m$ which is comparable to the thickness of the bubble wall in the previous example. 

Now, we wish to focus on a hybrid configuration in which the vortex line is piercing through a finite radius bubble. In polar coordinates, the ansatz for the Goldstone mode describing such a configuration is 
\begin{equation}
	\label{eq:ansatz02}
	\theta
		=
					n\.\phi
					+
					\omega\.t
					\, . 
\end{equation} 
The time-dependence of the Goldstone field guarantees stability of the bubble, whereas the winding number $n$ maintains the vortex within.

\begin{figure}[t]
	\centering
	\includegraphics[width = 0.42 \textwidth]
		{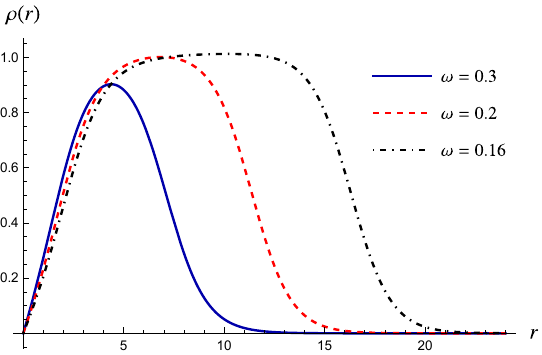}
	\vs{2mm}
	\caption{ Vortex profile in the $2-$dimensional case as a function of radius $r$ in units of $m$ for different charges.
	 Here, $m = f = 1$; values for $\omega$ are indicated in the legend.
	 }
	\label{fig:solution}
\end{figure}

As it is clear from Eqs.~(\ref{eq:thinwall}, \ref{eq:ERNG}), in the thin-wall limit, the radius of the bubble is much larger than the core of the vortex as well as the thickness of the bubble wall, $R \gg \delta_{\vrm} \sim \delta_{\wrm} \sim 1 / m$ (see Fig.~\ref{fig:solution} for an illustrative numerical solution in the cylindrical case).

\begin{figure}[t]
	\centering
	\includegraphics[width = 0.32 \textwidth]{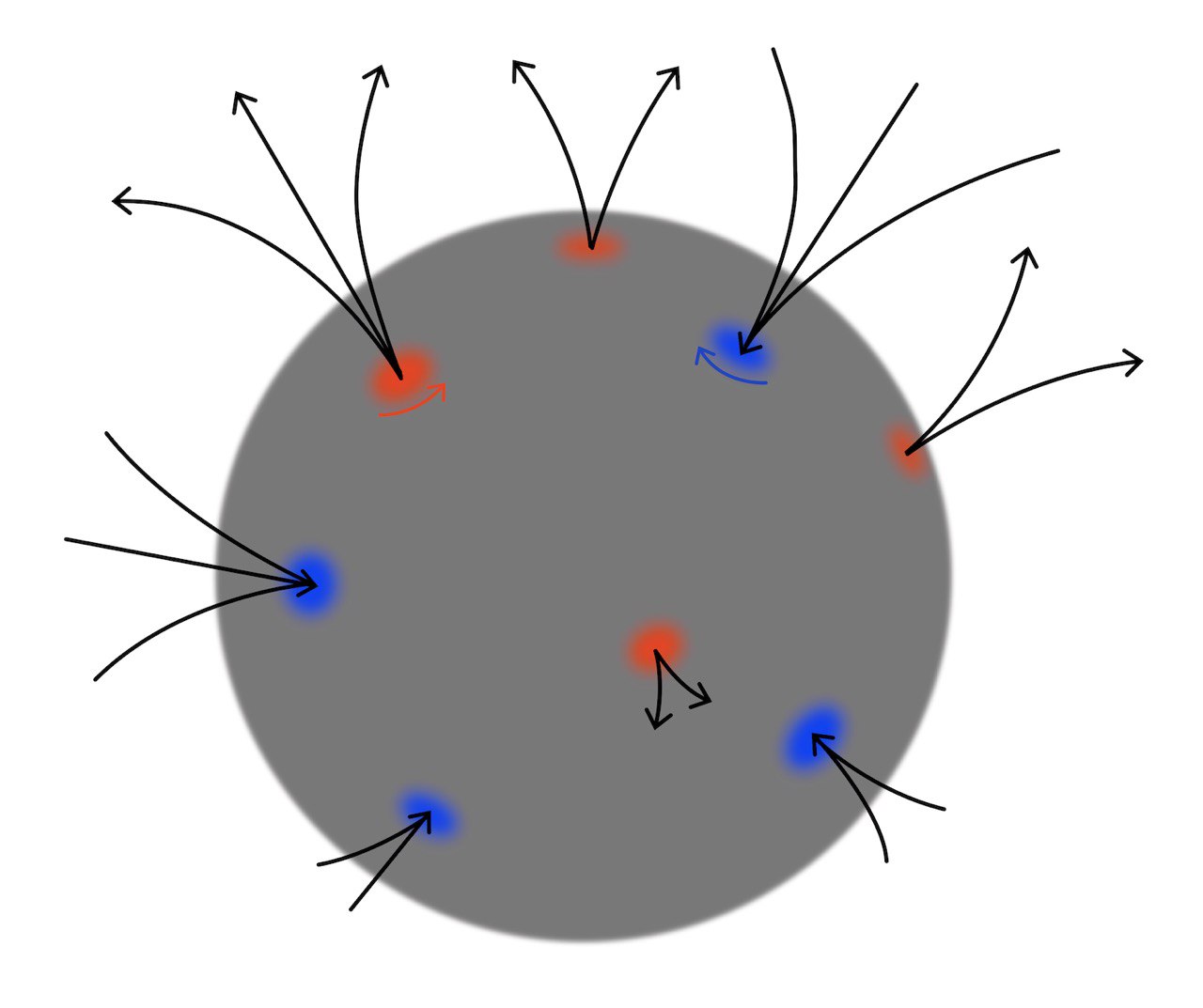}
	\caption{Sketch of a black hole with a number of 
		randomly oriented vortex/anti-vortex pairs.\\[-10mm]}
	\label{fig:vortexes}
\end{figure}

The above winding configuration endows the bubble with angular momentum $J = n\.N_{\Grm}$. This is very similar to the case of a spinning $Q$-ball considered in Ref.~\cite{Kim:1992mm} for $2 + 1$ dimensions and generalized to $3 + 1$ in Ref.~\cite{Volkov:2002aj}. The important difference is that for a saturated bubble $N_{\Grm}$ is also equal to entropy. Therefore, for such a bubble we have 
\vs{-1mm}
\begin{equation} 
	J
		=
					n\.N_{\Grm}
		\sim
					n\.S
					\, .
\end{equation} 
Notice that for the saturon bubble, the total vorticity $n$ cannot be much higher than one. In the opposite case, the vortex energy $\sim n^{2} / ( \alpha\mspace{1mu}R )$ would exceed the entire mass of the bubble, which is impossible. This imposes the upper bound $n \sim 1$. Thus, we learn that the maximal angular momentum of the saturon bubble is
\begin{equation} 
	J_{\rm max}
		\sim
					\frac{ M_{\rm Bub}^{2} }{ f^{2} }
		\sim
					S
					\, . 
\end{equation}
Strikingly, this expression is similar to the one satisfied by the maximal angular momentum of a spinning black hole,
\vs{-1mm}
\begin{equation}
	\label{eq:JMaxBH}
	J_{\rm max}
		\sim 
					\frac{ M_{\rm BH}^{2} }{ M_{\rm P}^{2} }
		\sim
					S
					\, .
\end{equation}
The above similarity is remarkable in two ways. First, it gives yet another supporting evidence to the correspondence between the generic saturons and black holes. Secondly, it offers a microscopic explanation of the bound \eqref{eq:JMaxBH}: the maximal angular momentum of a black hole is restricted by the vorticity of the graviton condensate. In addition, there can exist non-stationary non-axisymmetric configurations with zero vorticity, such as, e.g., a vacuum bubble or a black hole pierced through by multiple vortex lines with zero total winding number and unrelated spin (for a sketch see Fig.~\ref{fig:vortexes}).

It should be made clear that we are not suggesting that a black hole cannot spin without vorticity. This is already clear from the fact that for a large black hole the angular moment can change almost continuously, whereas the vorticity is defined by a topological winding number. However, our point is that the maximal spin of a black hole is necessarily correlated with the maximal vorticity that a black hole can sustain. In this way, we provide a topological meaning to extremality. Of course, some non-stationary black holes, can carry vortices even if the total spin is zero. But such configurations will evolve in time even classically. Thus, a necessary correlation between vorticity and spin takes place only for highly-spinning black holes.

Notice that as a very interesting byproduct, our picture offers a microscopic explanation for the zero temperature of extremal black holes. Thus, an extremal black hole in our picture is a black hole that has the maximal possible spin for a given mass. It is obvious that such objects cannot evaporate. Evaporation is a process that leads to a gradual decrease of the mass. Since the spectrum is thermal, emission of a quantum of arbitrarily low energy is possible. But for a black hole of maximal vorticity, such a process is not possible since the winding number cannot change continuously due to its topological nature. Our picture thus gives a topological meaning to the stability of extremal black holes. Notice that small corrections to thermality due to finite $N$ cannot change this conclusion, since the jump in winding number requires the change of energy of order one in units of a black hole mass. Such a process has an exponentially suppressed probability and vanishes in the strict semi-classical limit which represents the correct reference point for comparison with known properties of extremal black holes.

Also notice, the degeneracy of the Goldstone vacuum, that is responsible for maximal microstate entropy, is in one-to-one correspondence with the degeneracy of the vortex configurations. This is because the vortices are formed by the same broken generators that account for the bubble degeneracy. This connection between vorticity and the microstate entropy gives an additional basis for our proposal that analogous vorticity must be supported by black holes.\\[-3mm]

{\textsl{Trapping of Magnetic Fields}\;---\;}Despite the fact that the order parameter responsible for the vortex carries no gauge charge when interacting with some charged matter, the vortex will trap the flux of the corresponding magnetic field. This is a very general phenomenon of trapping the gauge flux by a global vortex \cite{Dvali:1993sg}, which we shall apply to the present case. 
 
Let us represent the vortex field in form of a complex order parameter $\psi \equiv \rho \.\erm^{i\mspace{1.5mu}\theta}$. This field carries no gauge charge. Let us assume that the vortex order parameter interacts non-trivially with some fields, $\chi_{+} = \rho_{+}\.\erm^{i\mspace{1.5mu}\theta_{+}} $ and $\chi_{-} = \rho_{-}\.\erm^{i\mspace{1.5mu}\theta_{-}}$, carrying opposite charges $ = \pm\.q$ under some gauge U$( 1 )_{\rm gauge}$ symmetry. Its r{\^o}le can be played by ordinary electromagnetism, the weak isospin, or some hidden sector symmetry. For example, these two fields can impersonate the two oppositely-charged components (e.g., electrons and ions) of the neutral plasma interacting with a black hole or a $Q$-ball.

Correspondingly, we assign the non-zero expectation values, $\langle \rho_{-} \rangle^{2}$, $\langle \rho_{-} \rangle^{2} \neq 0$, which measure the properly-weighted number densities of particles. The interaction term must respect both the global shift symmetry, $\theta \rightarrow \theta + \alpha$ by a constant phase $\alpha$, as well as, the U$( 1 )_{\rm gauge}$ gauge symmetry. Without loss of generality, we can consider the following interaction in an effective many-body Lagrangian \footnote{Such interactions are easily implementable in SU$( N )$-saturon models with $\chi_{\pm}$ transforming in appropriate representations of SU$( N )$.},
\begin{equation}
	\label{eq:Int}
	\psi\.\chi_{-}\.\chi_{+}
	+
	\,{\rm h.c.}
		=
					2\.\rho\.\rho_{+}\.\rho_{-}
					\cos(
						\theta
						+
						\theta_{+}
						+ 
						\theta_{-}
					)
					\, .
\end{equation} 
Eq.~\eqref{eq:Int} fixes the transformation properties under the global shift symmetry as $\chi_{\pm} \rightarrow \chi_{\pm}\.\erm^{i\alpha}$. This transformation is orthogonal to U$( 1 )_{\rm gauge}$. Despite this, the vortex is trapping the magnetic field. 

We can prove this by following the steps of Ref.~\cite{Dvali:1993sg}. Let us analyse the field configuration far away from the vortex core. Assuming that the electromagnetic field strength is zero, the equation for the photon field reads
\begin{equation}
	\label{eq:ansatz1}
	A_{\mu}
		=
					\frac{ 1 }{ e\mspace{1mu}q }\.
					\frac{
						\langle
							\rho_{+}
						\rangle^{2}\.\partial_{\mu}\theta_{+}
						-
						\langle
							\rho_{-} 
						\rangle^{2}\.\partial_{\mu}\theta_{-} }
						{
						\langle
							\rho_{-}
						\rangle^{2}
						+
						\langle
							\rho_{+}
						\rangle^{2}
					}
					\, ,
\end{equation}
where $e$ is the gauge coupling. Integrating the above expression over a closed path around the vortex and using Stoke's theorem, we obtain the following magnetic flux conducted by the vortex, 
\begin{equation}
	\label{eq:Flux}
	{\rm Flux}
		=
					\oint \d x^{\mu} A_{\mu}
		=	
					\frac{ 2\pi }{ e\mspace{1mu}q }\!
					\left[
						n_+
						+
						n\.
						\frac{\langle \rho_{-} \rangle^{2}}{
							\langle \rho_{-} \rangle^{2}
							+
							\langle \rho_{+} \rangle^{2}
						}
					\right]
					, 
\end{equation} 
where $n_{\pm} = \frac{ 1 }{ 2\pi }\.\oint \d x^{\mu} \partial_{\mu} \theta_{\pm}$ are integers and we took into account that, regardless of the coefficient of the interaction energy \eqref{eq:Int}, its minimization demands $n_{+} + n_{-} = -\.n$. This is non-zero, unless the expectation values are adjusted to $n$ via extreme fine-tuning, which cannot happen throughout the accretion process. In particular, for $n=1$ such an adjustment is simply impossible.
 
Despite being formed by an electrically neutral order parameter, when interacting with a neutral medium with dynamical charged components, the vortex generically traps the magnetic flux. Notice, this trapping is very different from the magnetic flux supported by Abrikosov (or Nielsen-Olesen) vortex lines in superconductors. In these cases, the primary order parameter transforms under the gauge U$( 1 )_{\rm gauge}$. Due to this, the flux is quantized in units of $1 / ( e\mspace{1mu}q )$. In contrast, the magnetic flux \eqref{eq:Flux} trapped by a global vortex is fractional \cite{Dvali:1993sg}. 

The configuration of a black hole with a trapped magnetic flux is similar to a classical solution of a black hole pierced by a cosmic string. Such solutions are well known (\cf~Ref.~\cite{Achucarro:1995nu}). In the case of electromagnetic U$( 1 )$, the r{\^o}le of the flux is played by an ordinary magnetic field trapped in form of a vortex line.

In addition, the magnetic flux can have a stabilising effect against unwinding of a vortex in a way similar to how the gauge flux stabilizes the vortices in theories with topologically-trivial vacuum structure \cite{Vachaspati:1991dz}. This effect can be explicitly traced in the present case of a saturon with the vortex with fractional magnetic flux. Thus, astrophysical black holes are expected to exhibit vortex structure with the magnetic flux lines piercing through them.

{\textsl{Consequences of Vorticity}\;---\;}It is well known that highly-rotating black holes can emit very powerful jets and in turn spin down (see Ref.~\cite{McKinney:2012vh} for a recent review). One important mechanism powering these jets has been suggested by Blandford \& Znajek (BZ) \cite{Blandford:1977ds}. The core idea relies on the presence of a black hole magnetosphere, which by rotation winds up in the toroidal direction and thereby leads to emission of powerful jets. For these to be sustained, the usual assumption is the presence of magnetized accreting matter around the black hole. However, for the BZ mechanism to work, the conditions concerning strength, length scale and configuration of the magnetic field must be chosen properly. Furthermore, magneto-rotational instabilities provide another complication \cite{McKinney:2012vh}.

Our proposal provides a very different framework for understanding this dynamics. As discussed in the previous section, vortices trap the magnetic field by interacting with the surrounding neutral plasma of either ordinary or a weakly-charged dark matter. Therefore, highly-rotating black holes, can efficiently slow down due to the emission process introduced in Ref.~\cite{Blandford:1977ds}. Their total emitted power can be estimated as $P_{\rm BZ} \sim {\rm Flux}^{2}\.\Omega^{2}$.

Of course, the coherence of the resulting jets highly depends on the geometrical distribution of the magnetic flux. Assuming the simplest alignment with the rotational axis, the emitted power can span several orders of magnitude within the range of observational interest. Note that the maximal magnetic field (of arbitrary gauge group) that a black hole can sustain is $B \sim {\rm Flux} / R_{\grm}^{2} \lesssim M_{\Prm}\.R_{\grm}^{-1} \sim M_{\Prm}^{2} / \sqrt{N_{\rm gr}\,}$.

One of the intriguing examples is the jet power observed in ${\rm M87}$, $P_{\rm M87} \sim 10^{42 \div 44}\,{\rm erg\,s}^{-1}$ \cite{EventHorizonTelescope:2021srq}. The numerical simulations within the standard framework seem not to provide jets powerful enough to be compatible with the high end of the above interval \cite{EventHorizonTelescope:2021srq}.\footnote{Note that an accretion disk might influence the BZ emission.} Within our framework, the effect can potentially be accounted solely by the BZ mechanism due to vorticity. 

In particular, the flux \eqref{eq:Flux} can be macroscopic if the charge $q$ is extremely small. For example, in case of ordinary electrodynamics, particles with very small $q$ can be present in the form of a fluid of dark matter or even a stationary vacuum condensate.

For example, a light cold dark matter particle of mass $m_{\rm DM}$ and local energy density $\rho_{\rm local}$ in the black hole vicinity, would generate the following effective mass for the photon, $m_{\gamma} = e\mspace{1mu}q\.\sqrt{\rho_{\rm local}\,} / m_{\rm DM}$. For illustration, let us choose both Compton wavelengths to be comparable with the gravitational radius of ${\rm M87}$-type black hole, $m_{\gamma} \sim m_{\rm DM} \sim 10^{-18}\,$eV. This in particular will guarantee the localization of the entire magnetic flux within the black hole neighbourhood. Then, for $e\mspace{1mu}q \sim 10^{-39}$, (implying $\rho_{\rm local} \sim 10^{4}\,{\rm eV}^{4}$) the black hole would be endowed with a jet power compatible with observations. 
 
In general, a constraint to be taken into account comes from the Galactic magnetic field $B_{\rm galaxy}$. In fact, due to the vorticity of the latter, the photon mass on galactic scales is bounded by $m_{\gamma} < \sqrt{e\mspace{1mu}q\,B_{\rm galaxy}\,}$ \cite{Adelberger:2003qx}.
 
The above discussion shows how the black hole vorticity could provide a ``portal" into the sector of dark matter with minuscule electric charges. Curiously, in the previous example, the mass $m_{\rm DM}$ is not too far from the range of so-called ``fuzzy" dark matter \cite{Hu:2000ke, Hui:2016ltb}. However, at the level of the present discussion we are not attempting to establish a specific connection. 

A further implication of the presence of vortices is their impact on the evaporation dynamics of a near-extremal black hole. Due to Hawking emission these objects are expected to become extremal, at which point, as discussed previously, evaporation stops. The astrophysical consequences of this will be explored elsewhere.

Another consequence of vorticity is related to mergers of compact bodies with at least one being in the mass range around a solar mass. These are generally believed to involve neutron stars. However, as gravitational-wave observatories are not yet sensitive enough to sufficiently resolve the information on the compactness of these objects, their classification is merely by mass, plus a possible electromagnetic counterpart \cite{Margutti:2017cjl}. While the former can be accounted for by a primordial nature of the hole, the latter could be explained by its vorticity.\\[-2mm]

{\textsl{Acknowledgments}\;---\;}We are thankful to Lasha Berezhiani and Andrei Gruzinov for useful discussions. Special thanks are to Roger Blandford for correspondence on black hole simulations. This work was supported in part by the Humboldt Foundation under Humboldt Professorship Award, by the Deutsche Forschungsgemeinschaft (DFG, German Research Foundation) under Germany's Excellence Strategy - EXC-2111 - 390814868, and Germany's Excellence Strategy under Excellence Cluster Origins.

\bibliography{refs}

\end{document}